\documentclass[prb,superscriptaddress,aps,twocolumn,10pt,showkeys]{revtex4-1}
\usepackage{graphics,graphicx,epsfig,color,latexsym,float,amsmath,bm}
\usepackage[colorlinks,citecolor=blue,linkcolor=blue,urlcolor=blue]{hyperref}
\usepackage[version=4]{mhchem}
\usepackage{wrapfig}

\usepackage{ulem}
\usepackage{lipsum}
\usepackage{placeins}
	
\newcommand{\avg}[1]{\left< #1 \right>} 
\newcommand{\ket}[1]{\left| #1 \right>} 

\begin{document}
\title{Skyrmion-driven topological Hall effect in a Shastry-Sutherland magnet}
\author{N. Swain}
\affiliation{School of Physical and Mathematical Sciences, Nanyang Technological University 637371, Singapore}
\author{M. Shahzad}
\affiliation{Department of Physics and Atmospheric Science, Dalhousie University, Halifax, Nova Scotia, Canada, B3H 4J5}
\author{G.~V. Paradezhenko}
\affiliation{Skolkovo Institute of Science and Technology, Moscow 121205, Russia}
\author{A.~A. Pervishko}
\affiliation{Skolkovo Institute of Science and Technology, Moscow 121205, Russia}
\author{D. Yudin}
\affiliation{Skolkovo Institute of Science and Technology, Moscow 121205, Russia}
\author{P. Sengupta}
\affiliation{School of Physical and Mathematical Sciences, Nanyang Technological University 637371, Singapore}
	
\date{\today}

\begin{abstract}
The Shastry-Sutherland model and its generalizations have been shown to capture emergent complex magnetic properties from geometric frustration in several quasi-two-dimensional quantum magnets. Using an $sd$ exchange model, we show here that {\it metallic} Shastry-Sutherland magnets can exhibit a topological Hall effect driven by magnetic skyrmions under realistic conditions. The magnetic properties are modeled with competing symmetric Heisenberg and asymmetric Dzyaloshinskii-Moriya exchange interactions, while a coupling between the spins of the itinerant electrons and the localized moments describes the magnetotransport behavior. Our results, employing complementary Monte Carlo simulations and a novel machine learning analysis to investigate the magnetic phases, provide evidence for field-driven skyrmion crystal formation for an extended range of Hamiltonian parameters. By constructing an effective tight-binding model of conduction electrons coupled to the skyrmion lattice, we clearly demonstrate the appearance of the topological Hall effect. We further elaborate on the effects of finite temperatures on both magnetic and magnetotransport properties.
\end{abstract}

\maketitle

\section{Introduction} 
With the experimental observation of a skyrmion lattice in three-dimensional helical magnets MnSi~\cite{Muhlbauer2009}, Fe$_{1-x}$Co$_x$Si~\cite{Yu2010}, and thin films of FeGe~\cite{Yu2011},  significant attention has been paid to the systematic study of noncollinear magnetic states and associated magnetoelectric phenomena, where magnetic skyrmions play a prominent role~\cite{Bogdanov1989,Bogdanov2001,Rossler2006,Nagaosa2013,Yudin2017}. Analogous to a topologically protected field-theoretical solution due to Skyrme~\cite{Skyrme1961,Skyrme1962} the stability of this particle-like spin texture is guaranteed by topological arguments, making it immune to smooth perturbations~\cite{Bogdanov1995,Pereiro2014,Hagemeister2015,Koumpouras2016}. In chiral magnets, skyrmions are stabilized due to a delicate interplay between direct exchange, favoring collinear ordering, and asymmetric exchange known as Dzyaloshinskii-Moriya interaction (DMI) that results in spin canting \cite{Bogdanov1989,Bogdanov2001}. Theoretical studies, including Monte Carlo (MC) simulations, based on two-dimensional classical spin models unambiguously reveal the formation of a skyrmion crystal in these systems~\cite{Yi2009,Han2010,Ambrose2013,Gungordu2016,Bottcher2018,Nishikawa2019,Mohanta2019}.

When it comes to emergent electrodynamics, a spin-polarized current flowing in a helical magnet is predicted to induce a torque that would produce new kinds of magnetic excitations~\cite{Schulz2012}. Under strong magnetoelectric coupling charge carriers moving through a noncoplanar magnetic texture are known to accumulate a finite Berry phase and were shown to exhibit the topological Hall effect~\cite{Lee2009,Neubauer2009,Kanazawa2011,Li2013,Yin2015,Ndiaye2017,Tome2021}. The propagating electrons experience a torque that results in alignment of their spins with the local moment~\cite{Hamamoto2015,Gobel2017,Denisov2017,Gobel2018,Gobel2019}. The gauge flux serves as an effective magnetic field acting on the electrons and maps the interaction of the electron {\it spin} and local moments to a magnetic field coupled to the {\it charge} of the itinerant electrons, analogous to quantum Hall systems on a lattice. The resulting Lorentz force drives a transverse current and can in its turn be used to detect the presence of skyrmions~\cite{Schulz2012,Gobel2019}.

In this paper, we address the formation of a magnetic skyrmion crystal within the Heisenberg model of classical spins on the Shastry-Sutherland lattice (SSL)~\cite{Shastry1981,Miyahara2003}. For this, we use large-scale MC simulations based on the \textsc{Metropolis} algorithm complemented by the machine learning (ML) approach as implemented in Ref.~\cite{Kwon2019} to determine the magnetic phase diagram. In the following, we consider a tight-binding model of the itinerant electrons on a skyrmion crystal to probe the topological Hall effect and subsequently examine its dependence with respect to the temperature, chemical potential, and applied magnetic field.

\section{Model}
\begin{figure*}[t]
\centering
\includegraphics[scale=1]{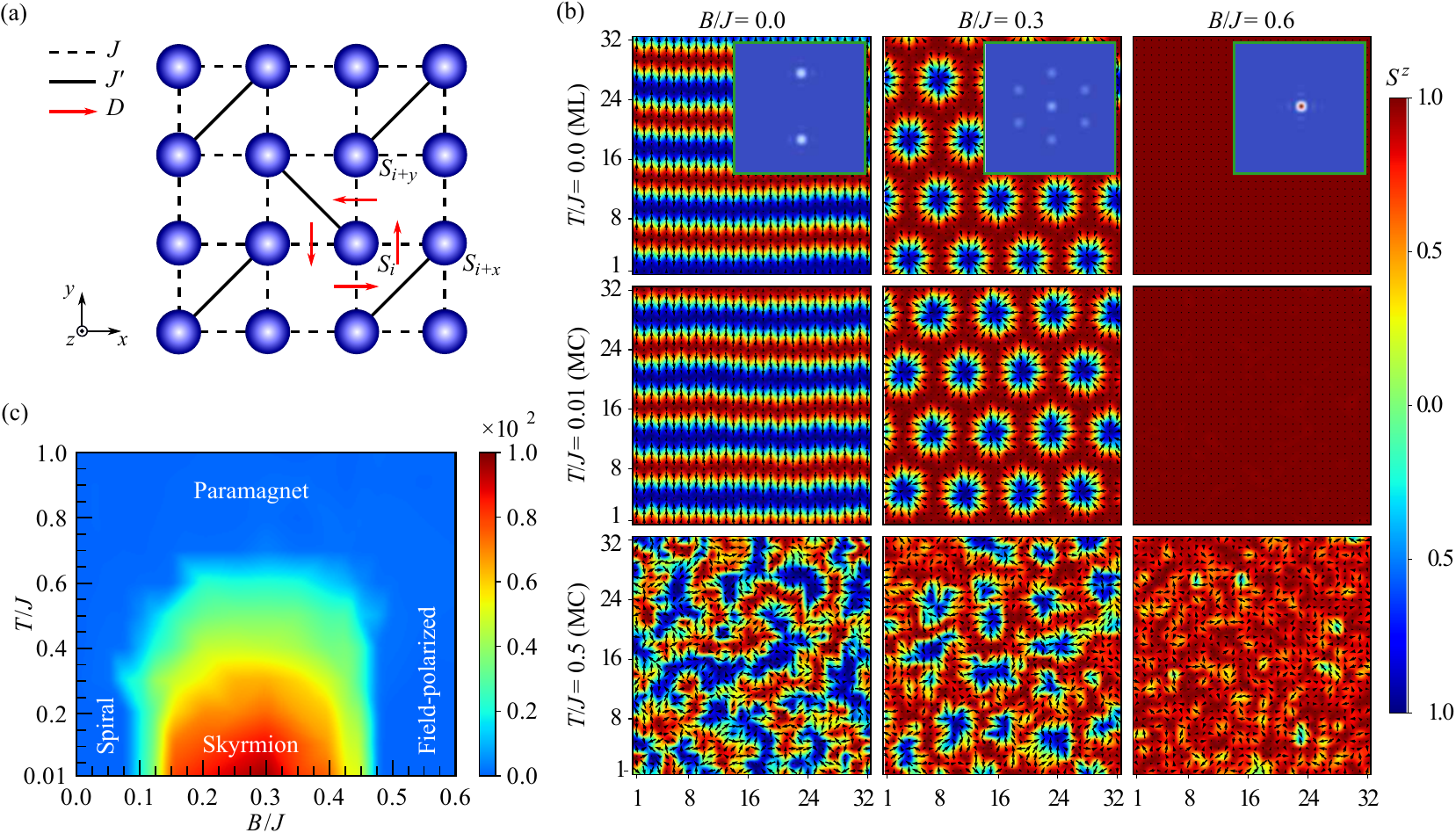}
\caption{\label{fig:phases_ssl}(a) A schematic of a two-dimensional magnet on the Shastry--Sutherland lattice (SSL). The localized magnetic moments are described by the classical Heisenberg Hamiltonian (\ref{d_ham}) that includes direct exchange of the coupling strength $J$ along the edges (dashed lines) and $J^\prime$ along diagonals (solid lines), as well as the Dzyaloshinskii-Moriya interaction (DMI) as shown by arrows. (b) Real-space spin textures are obtained by varying the external magnetic field, $B$, and temperature, $T$, based on machine learning optimization (ML) and Monte Carlo simulations (MC) at fixed $J'=-J$ and $D=0.5J$. The in-plane $(S_i^x,S_i^y)$ components of localized magnetic moments are depicted by black arrows in the $xy$ plane, whereas the out-of-plane $S_i^z$ components are visualized with the color bar. In the insets, corresponding static spin structure factors ${\cal S}(\bf{Q})$~(see Appendix~\ref{app:appendixa} for details) are shown that can be used to probe magnetic long-range ordering. Indeed, one peak in field-polarized phase, two peaks in a spin-spiral configuration, and seven peaks in a skyrmion crystal are clearly distinguishable. (c) The emergent spin chirality shown in terms of the external magnetic field and temperature evaluated from MC simulations. In full agreement with real-space configurations, chirality is zero in a spin-spiral and field-polarized states, whereas it is finite in a skyrmion crystal. Computational details of finite-temperature calculations are included in Appendix~\ref{app:appendixb}}.
\end{figure*}

We base our analysis on an $sd$-type exchange model of a two-dimensional magnet on the SSL, as schematically depicted in Fig.~\ref{fig:phases_ssl}(a). This model is defined by the Hamiltonian,
\begin{equation}\label{tot_ham}
    H=-t\sum_{\avg{i,j}} c_i^\dagger c_j-\mu\sum\limits_ic_i^\dagger c_i-J_K\sum\limits_i\bm{S}_i\cdot\bm{s}_i+H_d,
\end{equation}
with $J_K$ specifying the strength of the interaction between localized magnetic moments $\bm{S}_i$ and spins of conduction electrons $\bm{s}_i=c_i^\dagger\bm{\sigma}c_i$, linked to a vector of Pauli matrices $\bm{\sigma}=(\sigma_x,\sigma_y,\sigma_z)$. In Eq.~(\ref{tot_ham}), the summation over nearest neighbors $\avg{i,j}$ is assumed; the parameter $t$ stands for the hopping energy, $\mu$ corresponds to the chemical potential, and $c_i=(c_{i\uparrow},c_{i\downarrow})$ is the annihilation operator of the conduction electrons at the site $\bm{r}_i$. 

The localized magnetic moments are approximated by classical spins  with competing symmetric and asymmetric pairwise couplings,
\begin{equation}\label{d_ham}
    H_d = -\sum_{\avg{i,j}} J_{ij} {\bm S}_i \cdot {\bm S}_j - \sum_{\avg{i,j}} {\bm D}_{ij} \cdot ({\bm S}_i \times {\bm S}_j) -B \sum_i S_i^z,
\end{equation}
where the first term represents the standard Heisenberg exchange, with $J_{ij} = J$ along the edges and $J_{ij} = J'$ along the diagonal bonds; see Fig.~\ref{fig:phases_ssl}(a). The second term defines the Dzyaloshinskii-Moriya interaction specified by $\bm{D}_{ij} = D({\hat{\bm{z}}}\times\bm{r}_{ij})$, where $D$ is the corresponding coupling strength and a unit vector $\bm{r}_{ij}$ connects $i$th and $j$th sites~\cite{Yu2019}. The last term is the Zeeman coupling to an external magnetic field $B$ aligned with $\hat{z}$ axis. In the following, we assume that $\bm{S}_i$ is a classical vector field of unit length, $\vert\bm{S}_i\vert=1$.

\section{Methodology}
To explore the magnetic phases of the Hamiltonian~\eqref{d_ham} and associated real-space magnetic configurations, we used the standard machinery of classical MC based on the \textsc{Metropolis} algorithm. We simulate a finite-sized SSL of dimensions $L\times L$ ($L=32$, 40, and 48) with periodic boundary conditions. Efficient thermalization is ensured by simulated annealing~\cite{Moliner2009,Slavin2011,Shahzad2017,Shahzad2020} where the simulation is started from a random spin configuration corresponding to high temperature $T_\mathrm{high} \sim 2J$, followed by a reduction of temperature in steps of $\Delta T =0.01J$ until the lowest temperature $T_\mathrm{low} = 0.01J$ is reached. At each temperature, we use $5 \times 10^5$ MC sweeps for equilibration, and $5 \times 10^5$ MC sweeps (in steps of 5000 MC sweeps) for the measurement of physical observable. Metastable states at the low-temperature regime, near the phase boundaries are avoided by starting the simulation from a variational ground state and then increasing the temperature using the MC scheme.

The MC results for magnetic ground-state phases are supplemented by a novel machine learning (ML) optimization approach
that was recently proposed to probe the ground-state spin configurations in Ref.~\cite{Kwon2019}. Notably, the technique has not been applied to geometrically frustrated systems, such as SSL, so far. Followed by a brief description of the method, we make use of the approach to validate MC simulations. The ML method consists of training a fully connected neural network with no hidden layers for searching the ground states of magnetic systems. On each iteration of the training loop, we generate a batch $\bm{X}$ of size $n_{\mathrm{b}}$ of $n$-dimensional standard normal random vectors and feed it to the neural network. The input features are then decoded using formula $\bm{Y} = \bm{XW} + \bm{C}$, where $\bm{W}$ is the $(n \times 3m)$-dimensional weights matrix, $\bm{C}$ is the $(1 \times 3m)$-dimensional bias and $m = L \times L$ is the total number of lattice sites. The $(n_b \times 3m)$-dimensional matrix $\bm{Y}$ of output features is then reshaped to form a batch of size $n_{\mathrm{b}}$ of $m$ three-dimensional spins configurations $(\bm{S}_1,\ldots,\bm{S}_m)$ on a lattice, while the normalization of output spins $\bm{S}_i$ to unit vectors serves as an activation function. 

Next, we use the output spins for calculating the batch-averaged energy density that represents a penalty function to be minimized, 
\begin{equation}\label{cost}
    \langle \varepsilon \rangle_{n_b}
    = -\frac1m \sum_{i=1}^m \langle \bm{S}_i \cdot \bm{h}^{\rm eff}_i \rangle_{n_b},
\end{equation}
where $\bm{h}^{\rm eff}_i = -\partial H/\partial \bm{S}_i$ is the effective field determined by the model Hamiltonian~\eqref{d_ham}. The effective field is calculated using two-dimensional convolution of three-dimensional spins configuration on a square $L\times L$ lattice with the $(3\times 3\times 3\times 3)$-dimensional kernel that describes all the interactions of the spin $\bm{S}_i$ with its nearest and next-nearest neighbors. The neural network weights are trained to minimize \eqref{cost}. In our simulations, we set the hyperparameters equal to $n_{\mathrm{b}} = 1024$ and $n=64$, and we perform $10^5$ steps of the Adam optimizer~\cite{Adam2014} for training the weights with the learning rate $\alpha = 10^{-4}$. Upon completion of the training process, the neural network maps input features into the ground-state spins configuration. We implement the method using \textsc{TensorFlow} library for ML~\cite{TF} and perform our calculations on GPU.

\section{Magnetic phase diagram}\label{sec:phases}
Real-space magnetic configurations obtained by both MC simulations and ML optimization are shown in Fig.~\ref{fig:phases_ssl}(b). The interplay between symmetric and asymmetric exchange interactions results in a spin-spiral configuration being stabilized at $B=0$. Using standard nomenclature, this is a  1-{\bf{\it Q}} phase, where the static spin structure factor, ${\cal S}({\bf Q})$, in the inset of Fig.~\ref{fig:phases_ssl}(b) shows prominent peaks at the symmetry-related momenta, ${\bf Q}_{\text{sp}}=(0,\pm\pi/2)$. Upon increasing the external magnetic field this state evolves to a skyrmion lattice for $ 0.075J<B<0.45J$, where ${\cal S}({\bf Q})$ exhibits the characteristic six peaks at ${\bf Q}_{\text{Sk}}=(0,\pm\pi/2), (\pm\pi/4,\pm\pi/4)$, underscoring its 3-{\bf{\it Q}} nature [the additional peak at ${\bf Q}=(0,0)$ is a trivial one reflecting the finite net magnetization due to the applied field]. Finally, at high fields ($B\geq 0.45J$), the ground state evolves  to a fully spin-polarized state. We note that the ground-state spin configurations from both MC and ML approaches display excellent agreement, thus highlighting the accuracy of the new ML approach. Raising the temperature randomizes spin orientation owing to thermal fluctuations by weakening and eventually suppressing the magnetic order. As there is no in-plane anisotropy in our model, the configurations shown in Fig.~\ref{fig:phases_ssl}(b) have a companion configuration of the same energy that can be obtained by rotating all the spins by $\pi/2$ about the $\hat{z}$ axis.

Further insight into the nature of magnetic ordering is obtained from the static spin chirality, which measures the degree of noncoplanarity of the spin textures~\cite{Ishizuka2018} (see Appendix~\ref{app:appendixa} for more details). The low-field spiral phase represents a noncollinear though coplanar spin configuration with vanishing spin chirality, shown in Fig.~\ref{fig:phases_ssl}(c), that is dominated by thermal fluctuations. Once a skyrmion crystal is stabilized at $0.075J<B<0.45J$, the spin chirality sharply increases to a larger value, confirming its noncoplanar ordering. The spin chirality remains almost unchanged in a skyrmion crystal. With increasing temperature, the skyrmion crystal melts into a skyrmion liquid with a corresponding reduction in the chirality. Finally, at high temperatures, the chirality vanishes completely in the paramagnetic phase. At higher magnetic fields, the increase of spin polarization results in spin chirality going down. Eventually, spin chirality drops to zero at $B/J \geq 0.45$, since the spins are aligned with the direction of a magnetic field and noncoplanarity is lost. Note that the results of our numerical simulations are presented for $J' = -J$ and $D = 0.5J$. A qualitatively similar behavior was observed for different values of $J$, $J^\prime$, and $D$ characterizing the exchange coupling strength along the edges and diagonals of SSL, as well as DMI, respectively.

\begin{figure*}[ht!]
\centering
\includegraphics[width=1.0\textwidth]{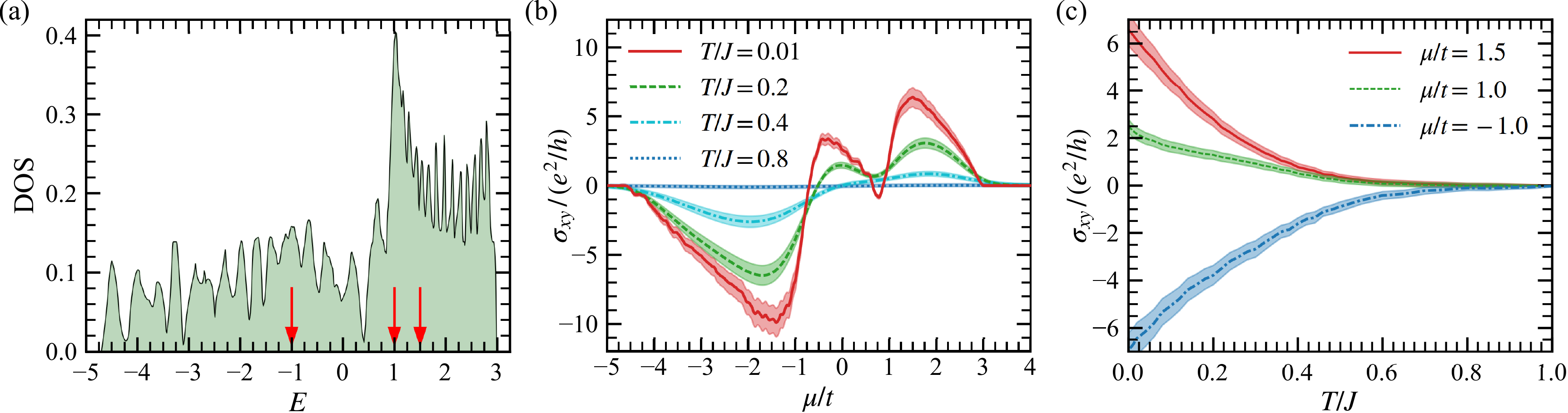}
\caption{\label{fig:cond}(a) Density of states of the itinerant electrons coupled to a skyrmion crystal as specified by the Hamiltonian (\ref{tot_ham}). Note that the hopping parameters in Eq.~(\ref{tot_ham}) along the edges and diagonals are chosen to be equal, and the energy $E$ is measured in the units of $t$. The transverse conductivity $\sigma_{xy}$, evaluated with the Kubo formula (\ref{conductivity2}) at $B/J = 0.3$, features the topological Hall effect in a noncoplanar magnetic background on varying the chemical potential (b), and temperature (c) with the red vertical arrows in (a) being a guide for the eye. Lines represent the conductivity averaged over multiple MC simulations, while the shaded areas represent the corresponding errors.}
\end{figure*}

\section{Topological Hall effect} 
In metallic magnets, the electronic transport properties are strongly modified by coupling to the underlying magnetic ordering. For simplicity, we consider a one-band model of itinerant electrons coupled to the local moments via an $sd$-type exchange interaction as specified by the Hamiltonian (\ref{tot_ham}). In practice, the latter can be realized in TmB$_4$ that represents alternating-layer crystalline structure, where the planes of thulium atoms are positioned in the middle between the planes of boron atoms. Interestingly, the sublattice of Tm atoms can be equivalently represented by the Shastry--Sutherland model with perfect squares and almost equilateral triangles, such that localized electrons with a large magnetic moments interact with conduction electrons in boron-derived bands~\cite{Shin2017}. We focus on the skyrmion crystal where the effect on magneto-transport is most dramatic.  In magnetic metals, in addition to the ordinary Hall resistance that is proportional to the applied magnetic field, anomalous and topological terms are present. The anomalous Hall effect appears in metals with a net magnetization due to spin-orbit coupling, while the topological Hall effect (THE) arises due to the real-space Berry phase that an electron moving through a noncoplanar spin texture acquires. In the following, we focus on the contribution to the transverse conductivity due to a skyrmion crystal, that exhibits THE due to finite chirality and resulting in the Berry phase. In contrast, the chirality vanishes in the spiral and fully polarized phases and hence no THE is observed. In addition to the zero-temperature transport, we also explore the effects of finite temperature on the THE. 

In the strong coupling limit $J_K\gg t$, the effective magnetic field produced by the magnetic ordering couples directly to the charge of the electron, analogous to quantum Hall systems~\cite{Hamamoto2015,Gobel2017,Gobel2018,Gobel2019}. The electronic properties can be described by an effective tight-binding model with a complex hopping matrix element whose phase depends on the relative orientation of the local moments as discussed in Appendix~\ref{app:appendixc}. The energy spectrum consists of bands grouped into a low-energy and high-energy sector with spins aligned parallel and antiparallel to the spins in a skyrmion, respectively, and separated by energy $\sim J_K$. The distribution of the bands depends on the degree of frustration, $t'/t$ (see Appendix~\ref{app:appendixc} for details). For the results shown here, we chose $t'/t=1$, in accordance with the ratio of the corresponding Heisenberg exchange interactions, $J'/J$. The bands are gapped and carry a finite Chern numbers. The resulting density of states (DOS) is shown in Fig.~\ref{fig:cond}(a)---the sequence of the minima confirms the presence of finite gaps between successive bands in the spectrum. 

To determine the Hall conductivity, we use the Kubo formula
\begin{equation}\label{conductivity2}
    \sigma_{xy}=\frac{ie^2\hbar}{N}\sum_{\sigma}\sum_{m\neq n}
    \frac{f_n-f_m}{E_m-E_n}\frac{\langle n\vert v_x\vert m\rangle\langle m\vert v_y\vert n\rangle}{E_n-E_m+i\eta},
\end{equation}
where indices $m$ and $n$ represent the sum over the energy levels, $N$ is the total number of sites, $f_k$ denotes the Fermi-Dirac distribution for the energy $E_k$, specified by the single-particle state $\ket{k}$. Note that the small broadening, $\eta$, introduced to Eq.~(\ref{conductivity2}), is associated with the conduction electrons scattering off the localized magnetic moments. The matrix elements of the velocity operators $\hat{v}_x$ and $\hat{v}_y$ are obtained using the equilibrium MC spin configuration. We diagonalize the effective tight-binding Hamiltonian ($N \times N$) to obtain its eigenspectrum and using Eq.~(\ref{conductivity2}), we calculate the Hall conductivity. We further use the ensemble of equilibrium MC configurations at a specific temperature and chemical potential to calculate the average Hall conductivity and its variance.

The behavior of the topological contribution to the transverse conductivity, $\sigma_{xy}$, with varying chemical potential in the skyrmion lattice state is shown in Fig.~\ref{fig:cond}(b) for three different temperatures, illustrating the effect of strong magneto-electric coupling ($\sigma_{xy}$ vanishes identically in the spiral and fully polarized states). At $T = 0.01J$, the transport properties are mainly determined by electron states below the Fermi energy. The noncoplanarity of the spin texture in a skyrmion crystal phase results in a finite Berry phase being acquired by the conduction electrons which shows up as a finite THE. Similar to the quantum Hall effect, in a skyrmion crystal phase, transverse conductivity exhibits a sequence of quantized plateaus corresponding to the filling of successive (gapped) electron energy bands with finite Chern numbers. With increasing chemical potential, the value of $\sigma_{xy}$ decreases from zero in steps of $e^2/\hbar$, in accordance with the Chern number $\mathcal{C} = -1$ of the low-energy sector. However, in contrast to quantum Hall systems with flat electron energy bands, one finds that the conductivity changes continuously, yet non-monotonically between the plateaus, which reflects the finite dispersion of the energy bands. At small, but finite temperatures, the chemical potential dependence retains its qualitative behavior, but the plateaus get washed out. Subsequent raising of the temperature leads to the randomization of the spins and vanishing of net chirality, which smears out the effect. The effect of temperature on THE is further illustrated in Fig.~\ref{fig:cond}(c) where we show the evolution of $\sigma_{xy}$ with temperature at fixed $\mu$. The magnitude of $\sigma_{xy}$ decreases monotonically with $T$ and vanishes in the paramagnetic phase. It should be noted that the quantization of the Hall plateaus is most pronounced for skyrmions with small radii where the lower bands are well separated. For larger skyrmions, the density of the bands increases and the energy extent of the conductivity plateaus decreases proportionately as does the energy separation between successive plateaus. This makes it difficult to resolve them in numerical simulations owing to finite-size effects and fluctuations.

\section{Conclusion}
In this paper, we discussed the THE on an SSL due to noncoplanar magnetic texture. We provided a detailed analysis of magnetic phase diagram for a Heisenberg model of classical spins based on MC simulations and ML optimization, including a proper treatment of temperature effects. Our numerical findings reveal the formation of a skyrmion crystal on an SSL, which, accompanied with the calculations of transport of conduction electrons, allow us to address the THE. Despite its simplicity, such a model makes it possible to capture relevant experimental signatures of the phenomenon~\cite{Soumyanarayanan2017,Raju2019}. We expect that our results will be crucial in understanding experimental observation and designing new experiments to realize topological magneto-transport properties in metallic Shastry-Sutherland magnets.

\section*{Acknowledgements}
We acknowledge the use of the computational resources at the High Performance Computing Centre (HPCC) at NTU (Singapore), the National Supercomputing Centre (NSCC) ASPIRE1 cluster (Singapore), and the Skoltech CDISE supercomputer ``Zhores''~\cite{Zhores} in our numerical simulations. The work of A.A.P. was supported by the Russian Science Foundation Project No.~20-72-00044. P.S. acknowledges support from the Ministry of Education (MOE), Singapore, in the form of AcRF Tier 2 Grant No. MOE2019-T2-2-119.

\appendix
\section{Characterizing magnetic ground-state phases}\label{app:appendixa}
\begin{figure}[b]
\centering
\includegraphics[width=0.45\textwidth]{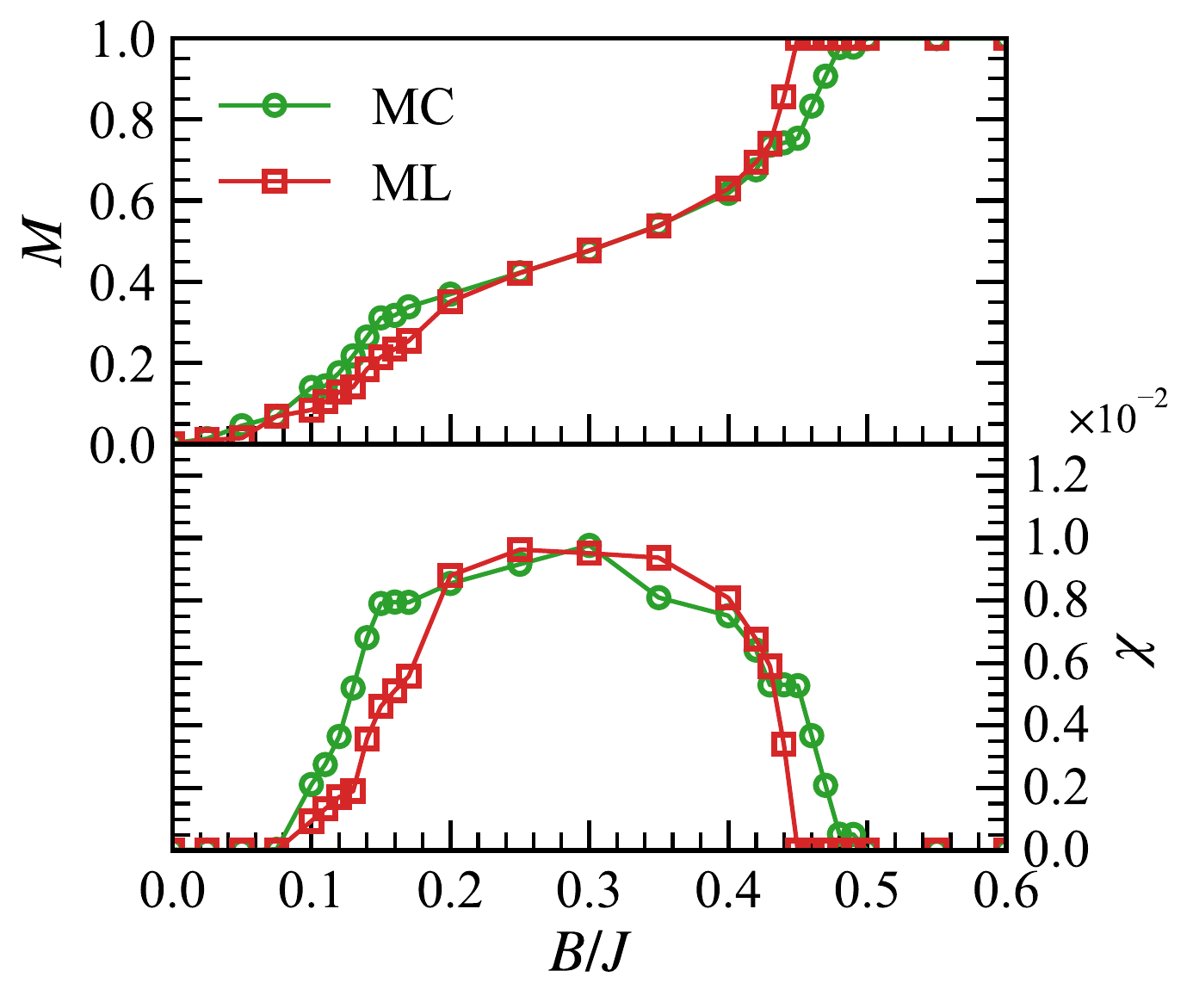}
\caption{\label{fig:mag_char}Magnetization $M$ and spin-chirality $\chi$ calculated from MC and ML spin configurations as a function of external magnetic field.}
\end{figure}

In Sec.~\ref{sec:phases}, to characterize the magnetic phases, we calculate the magnetization, spin chirality, and static spin structure factor. The net magnetization per site is defined as  
\begin{equation}
    M=\frac{1}{N} \langle\sum_{i} \bm{S}_i\rangle,
\end{equation}
where $N$ is the total number of spins on the SSL and $\avg{\ldots}$ is the average over multiple MC simulations or over the batch of size $n_b$ for ML optimization. The magnetization increases continuously with increasing the field from zero to full saturation ($B/J \sim 0.45$) as seen in the upper panel of Fig.~\ref{fig:mag_char}. However, the magnetization increase in the different phases exhibits distinct characteristics that distinguishes the phases. Importantly, the MC results for the lowest temperature match well with the ML optimization except close to the phase boundaries, underscoring the validity of the new approach.

\begin{figure*}[t]
\centerline{
	\includegraphics[width=0.8\textwidth]{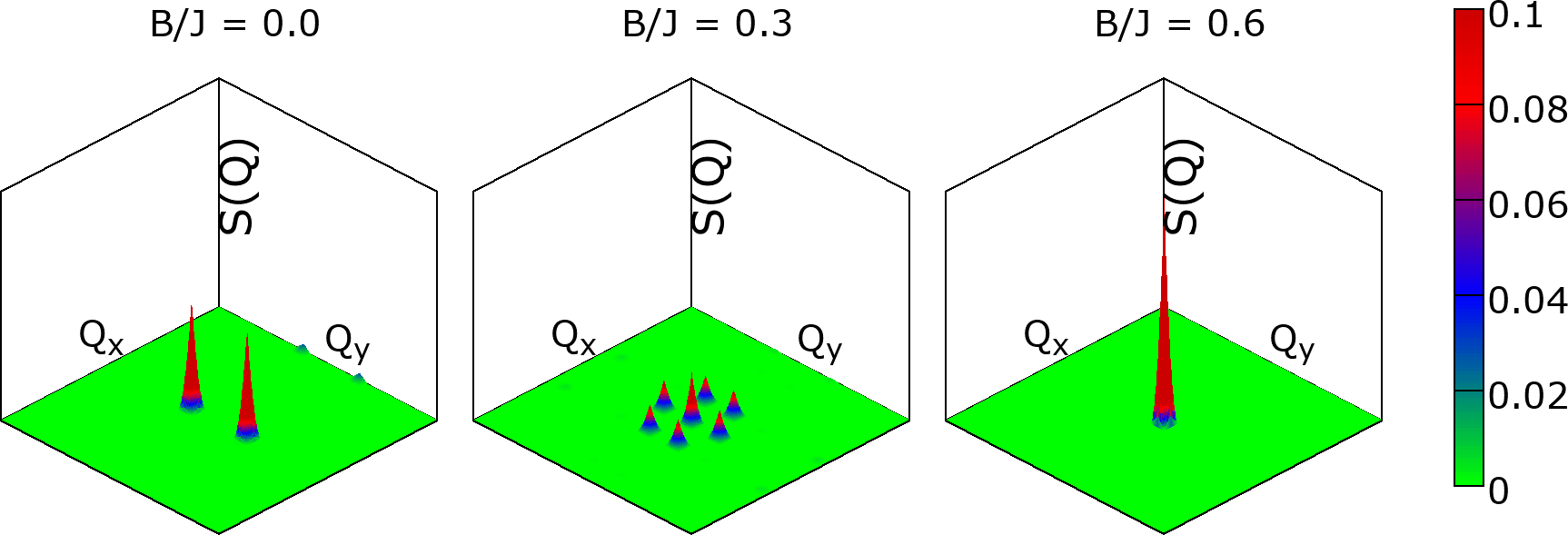}
}
\caption{Static spin structure factor at $B=0$ (a spin-spiral phase), $B=0.3J$ (a skyrmion crystal), and $B=0.6J$ (a field-polarized state).\label{fig:SF3D}
}
\end{figure*}


\begin{figure*}[ht!]
\centerline{
	\includegraphics[width=0.85\textwidth]{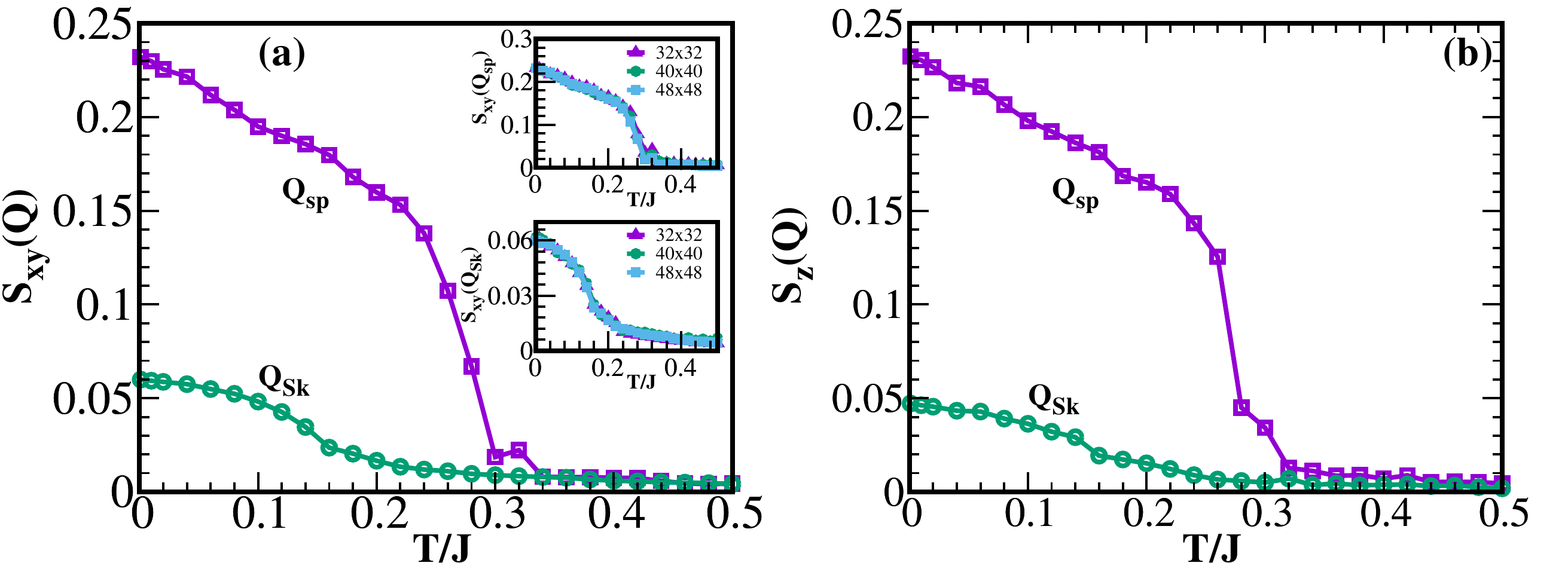}
}
\caption{\label{fig:finT}
Behavior of the peak height of the static spin structure factor components; $\mathcal{S}_{xy}({\bf Q})$ (transverse) and $S_{z}({\bf Q})$ (longitudinal) for different magnetic phases as a function of temperature.
}
\end{figure*}

The static spin chirality measures the degree of noncoplanarity of the spin configurations and is defined as 
\begin{equation*}
    \chi 
    = \frac{1}{4\pi N} 
    \langle \sum_{\Delta} \bm{S}_i \cdot (\bm{S}_{j} \times \bm{S}_{k}) \rangle,
\end{equation*}
where the triple product of the spins is taken over triangular plaquettes defined by the diagonal bonds. Chirality presents an important defining order parameter in the study of complex spin textures as it provides a direct indicator of noncoplanarity of the underlying magnetic order. For the present problem, $\chi = 0$ in the spin-spiral phase confirming its coplanar structure, and rapidly rises to a finite value at the field-driven transition to the skyrmion crystal phase, reflecting the noncoplanarity of the magnetic texture. It should be noted that the chirality remains roughly constant in the skyrmion phase for different magnetic fields, whereas at $B/J \geq 0.45$, chirality goes down to zero in the collinear field-polarized state. The results from both MC and ML simulations for $\chi$ with varying $B$ are shown in the lower panel of Fig.~\ref{fig:mag_char}. As with the magnetization, the data from the two approaches agree well except in the vicinity of the phase boundaries.

The static spin structure factor is defined as the Fourier transform of the spin-spin correlation,
\begin{equation*}
  \mathcal{S}({\bf{Q}})
  = \frac{1}{N^2} \sum_{i,j} 
  \avg {{\bm {S}}_i \cdot {\bm{S}}_j} 
  e^{i{\bf{Q}} \cdot ({\bm{r}}_{i} - {\bm{r}}_{j})},
\end{equation*}
where ${\bm {r}}_{i}$ specifies the position of the $i$th lattice site. The spin structure factor quantifies long-range magnetic order via peaks in the momentum space $\mathbf{Q}=(Q_x,Q_y)$, shown in Fig.~\ref{fig:SF3D}.

When the magnetic field is relatively weak the system is in the noncollinear spin-spiral state; $\mathcal{S}({\bf{Q}})$ has two distinguishable peaks that are related by symmetry and hence the state is identified by a single wave vector (1-$Q$ state). Increasing the field strength results in a transition to the skyrmion crystal phase; $\mathcal{S}(\bf{Q})$ exhibits the distinct six-peak structure characteristic of the skyrmion state. The six peaks represent three sets of symmetry-related pairs and the Skyrmion phase can be considered as a linear superposition of three spiral phases (3-$Q$ state). There is a trivial peak at ${\bf Q}=(0,0)$ that reflects a net magnetization due to the applied field. At stronger magnetic fields, the system has a tendency towards the formation of spin-polarized ferromagnetic phase; $\mathcal{S}(\bf{Q})$ has only one peak at ${\bf Q}=0$.

\begin{figure*}[t]
\centerline{
	\includegraphics[width=0.9\textwidth]{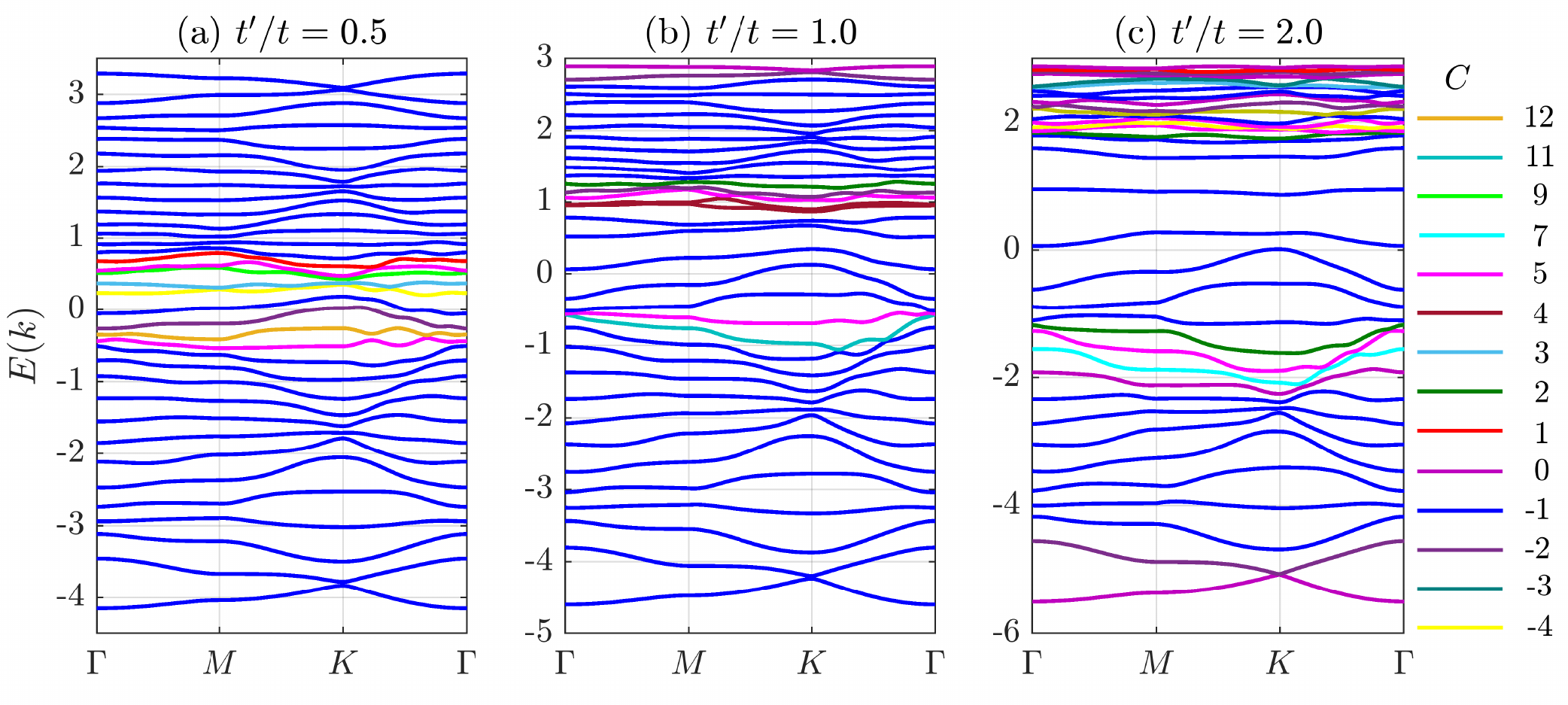}
}
\caption{\label{fig:band-str}
Band structure of the conduction electrons on a skyrmion crystal background. The bands, drawn in blue, have Chern number $\mathcal{C}=-1$, whereas the rest are marked with different colors (the colorbar to the right shows their particular values).
}
\end{figure*}

\section{Finite-temperature transitions}\label{app:appendixb}
Both the phases of interest, a spin-spiral and skyrmion crystal, extend to finite temperatures. With increasing temperatures, the spins are randomized by thermal fluctuations, which eventually drive a transition to the paramagnetic phase, which can be probed through the finite-temperature MC simulations. The transition to the paramagnetic phase is most conveniently studied through the evolution of the peak height in static structure factors in the respective phases. Figure~\ref{fig:finT} shows the simulation results for the temperature dependence of the height of the structure factor peak in the spiral phase and one of the peaks in the skyrmion crystal phase. Data for both the transverse, $\mathcal{S}_{xy}({\bf Q})$, and longitudinal, $\mathcal{S}_z({\bf Q})$, components of the static structure factor are shown as they yield complementary information. The strength of the structure factor peak decreases and eventually vanishes reflecting the suppression of long-range order. For the spiral phase, there is a direct transition to the paramagnetic phase. The skyrmion crystal, on the other hand, melts to form a skyrmion liquid. While the periodic arrangement of the skyrmions is lost, the individual skyrmions maintain their noncoplanar structure. This is reflected in the non-vanishing of the spin chirality in Fig.~\ref{fig:phases_ssl}(c). Thermal fluctuations reduce the noncoplanar ordering and  eventually at a sufficiently high temperature, the paramagnetic phase is reached. The inset shows the finite-size dependence of the structure factor peaks in both the phases and confirms the convergence of the data to the thermodynamic limit.

\section{Skyrmion band structure}\label{app:appendixc}
In the limit of $J_K \gg t$, the electron spins are completely aligned with the local moments and the Hamiltonian (\ref{tot_ham}) reduces to an effective tight-binding model,
\begin{equation}
 {\mathcal{H}_e}= -\sum_{\avg{i,j},\sigma}t_{ij}^\mathrm{eff}(d_{i}^\dagger d_{j}+\mathrm{H.c.}),
\end{equation}
where
\begin{equation}
t_{ij}^\mathrm{eff}=t e^{ia_{ij}}\cos\frac{\theta_{ij}}{2},
\label{eq:teff} 
\end{equation}
is the effective hopping matrix for the spin-parallel electrons between sites $i$ and $j$ and
\begin{equation}
a_{ij}=\arctan\frac{-\sin(\phi_i-\phi_j)}{\cos(\phi_i-\phi_j)+\cot\frac{\theta_i}{2}\cot\frac{\theta_j}{2}}
\end{equation}
is the phase factor, while $\theta_{ij}$ stands for the angle difference between two localized spins $\bm{S}_i$ and $\bm{S}_j$. The spin anti-parallel electrons are described by a similar effective tight-binding model with a different $t_{ij}^\mathrm{eff}$ and the two sectors are completely decoupled. The effective band structure is shown in Fig.~\ref{fig:band-str} for three representative values of the degree of frustration $t'/t$. We observe that the distribution of bands depends on this parameter, but share some common features. The bands are gapped and carry finite Chern numbers. In contrast to quantum Hall plateaus, the bands are not completely flat and have finite dispersion. Moreover, the Chern numbers for the bands are not restricted to $\pm 1$, but take multiple values, as indicated in Fig.~\ref{fig:band-str}. This is characteristic of a skyrmion crystal phase and results in the observed behavior of the transverse conductivity.

\FloatBarrier

\bibliographystyle{apsrev4-2}
\bibliography{main.bbl}

\end{document}